\documentclass{article}
\usepackage[T1]{fontenc} 
\usepackage[utf8]{inputenc} 
\usepackage{ismir,amsmath,cite,url}
\usepackage{graphicx}
\usepackage{color}
\usepackage{amsfonts}

\usepackage{booktabs}
\usepackage[inline]{enumitem}
\usepackage{xcolor}

\usepackage{lineno}

\newcommand{\ap}[1]{\textcolor{black}{#1}}
\newcommand{\mr}[1]{\textcolor{black}{#1}}
\newcommand{\bug}[1]{\textcolor{black}{#1}}

\title{From Discord to Harmony: Decomposed Consonance-based Training for Improved Audio Chord Estimation}

\multauthor
{Andrea Poltronieri \hspace{2cm} Xavier Serra \hspace{2cm} Martín Rocamora} {
  Music Technology Group, Universitat Pompeu Fabra\\
  {\tt \{andrea.poltronieri, xavier.serra, martin.rocamora\}@upf.edu}
}

\def\authorname{Andrea Poltronieri, Xavier Serra, Martín Rocamora}

\begin{document}

\maketitle
\begin{abstract}

Audio Chord Estimation (ACE) holds a pivotal role in music information research, having garnered attention for over two decades due to its relevance for music transcription and analysis. 
Despite notable advancements, challenges persist in the task, particularly concerning unique characteristics of harmonic content, which have resulted in existing systems' performances reaching a glass ceiling. 
These challenges include annotator subjectivity, where varying interpretations among annotators lead to inconsistencies, and class imbalance within chord datasets, where certain chord classes are over-represented compared to others, posing difficulties in model training and evaluation. 
As a first contribution, this paper presents an evaluation of inter-annotator agreement in chord annotations, using metrics that extend beyond traditional binary measures. 
In addition, we propose a consonance-informed distance metric that reflects the perceptual similarity between harmonic annotations.
Our analysis suggests that consonance-based distance metrics more effectively capture musically meaningful agreement between annotations. 
Expanding on these findings, we introduce a novel ACE conformer-based model that integrates consonance concepts into the model through consonance-based label smoothing.
The proposed model also addresses class imbalance by separately estimating root, bass, and all note activations, enabling the reconstruction of chord labels from decomposed outputs.



\end{abstract}


\section{Introduction}
\label{sec:intro}

In Western music theory, chords denote simultaneous combinations of three or more notes, forming harmonic structures integral to musical composition and analysis~\cite{dehaas2013similarity, deberardinis2023harmory, huang2014classification, pauwels2013segmentation}.  
However, manually annotating chords from audio recordings is a labour-intensive task requiring music professionals' expertise. 
Consequently, Audio Chord Estimation (ACE) emerged as a crucial task in Music Information Retrieval/Research (MIR) to automate chord transcription from audio due to its relevance for its numerous applications in music transcription and analysis. 

The research in ACE has witnessed more than two decades of exploration, but despite the important advancements achieved~\cite{pauwels2019twentyyears}, performance results have stagnated in recent years, leading some researchers to suggest that the task has hit a glass ceiling \cite{carsault2018loss-similarity}.
These challenges stem from several significant open problems~\cite{pauwels2019twentyyears}, which are fundamentally linked to the complex nature of harmonic content and its representation within audio signals.

One such challenge is the chord vocabulary imbalance, stemming from the unequal frequency of occurrence among chord labels. For instance, in \emph{ChoCo}~\cite{deberardinis2023choco}, the most extensive corpus of chord annotations to date, approximately $74.9\%$ of the distribution of the $8064$ distinct chord classes is dominated by just major, minor, major seventh, minor seventh and dominant seventh chord types.

Another critical challenge is inter-annotator agreement, which arises from the inherent ambiguity in what constitutes a chord from a musical perspective and the subjective nature of human annotation processes. 
For example, a clear distinction between a chord sequence and a melodic line can be subject to individual interpretation. Moreover, there is significant variance among annotators regarding the level of detail in annotating chord sequences~\cite{koops2019subjectivity}. 

Various studies have investigated inter-annotator agreement in chord annotation, reporting agreement rates for the root note ranging from $76\%$~\cite{koops2019subjectivity} to $92\%$~\cite{clercq2011rockcorpus}, using different datasets and numbers of annotators. 
Such evaluations typically use binary metrics to compare labels, but penalising the agreement evaluation equally for every discrepancy can be inappropriate~\cite{mcleod2023newmetrics}. 
Indeed, binary evaluation risks overlooking harmonic aspects that might be shared among chord sequences, although annotated differently.


As a preliminary contribution of this paper, we analyse patterns of inter-annotator disagreement in chord annotation. 
Our analysis reveals that when annotators disagree, their chord labels tend to be harmonically related rather than randomly different. 
Specifically, we find that disagreements commonly occur between chords that share significant harmonic content (c.f. Section \ref{sec:agreement}).

Building upon these insights, we propose a method for incorporating such information into the supervised training of ACE systems. 
Hence, we introduce a novel model integrating consonance-based label smoothing \cite{muller2019smoothing} (c.f. Section \ref{subsec:smoothing}).
To tackle the class imbalance issue, instead of mapping audio features to a predetermined vocabulary of chord labels, we adopt an approach inspired by \cite{mcfee2017cnn}, in which the chord root, bass, and all note activations are classified separately. 
The final predicted chord label is derived from decoding these three sets of information without explicitly imposing any vocabulary on it (c.f. Section \ref{subsec:vocabulary}).

The proposed model leverages the Conformer architecture \cite{gulati2020conformer}, which has 
\mr{recently been} explored 
\mr{in several music audio applications}~\cite{tamer2023violin, won2023foundation, Poltronieri2024ChordSync}.
We demonstrate that the proposed model performs better than the state-of-the-art approaches, especially when evaluated using non-binary and consonance-based distance metrics (c.f. Section \ref{sec:evaluation}).

\section{Related Work}
\label{sec:related}

Since Fujishima’s early work~\cite{fujishima1999ace}, chord recognition has followed knowledge-driven approaches~\cite{mcvicar2014acesurvey}, typically extracting chroma~\cite{mauch2010ace} or Tonnetz features~\cite{humphrey2012acetonnentz}, and classifying them via HMMs, DBNs~\cite{mauch2010ace}, or CRFs~\cite{korzeniowski2016crf}.


With the emergence of deep learning, various architectures have been explored for the task, including Convolutional Neural Networks (CNNs)~\cite{mcfee2017cnn, korzeniowski2016crf}, Recurrent architectures (RNN)~\cite{sigtia2015rnn}, Convolutional Recurrent Neural Networks (CRNNs)~\cite{jiang2019large-vocabulary-decomposition}, and Transformers~\cite{park2019bi-directionaltransformer}. 
While deep-learning approaches have surpassed traditional knowledge-driven ones, several challenges must be tackled. 
Most of the proposed approaches to addressing the chord class imbalance challenge can be divided into two categories: chord simplification and chord decomposition. The former reduces the size of the chord vocabulary by converting complex chord labels into simpler representations. 
Notably, the vast majority of studies have adopted restricted vocabularies of approximately 25 symbols, encompassing major-minor chords \cite{fujishima1999ace, mcvicar2014acesurvey}.
Chord decomposition strategies focus on predicting the chord constituting components separately, 
and then map them to templates to predict the final chord~\cite{mcfee2017cnn, wu2018acebigvocabularymidi, jiang2019large-vocabulary-decomposition}. 
Some additional approaches do not fall into these two categories, like addressing the unequal distribution of chords through a balanced learning process~\cite{deng2017aceevenchance}, or using a curriculum learning training scheme to begin with simple chord qualities and then move to more complex and less common ones~\cite{rowe2021acecurriculum}. 

The inter-annotator agreement in chord annotation continues to pose a significant challenge. 
Despite existing diagnoses and quantification of this phenomenon in the literature \cite{koops2019subjectivity, clercq2011rockcorpus}, definitive solutions have yet to emerge.
Clercq et al. \cite{clercq2011rockcorpus} observe an inter-annotator agreement rate of $94\%$ for the root note between two different annotations of the top 20 tracks from Rolling Stone magazine's list of the \emph{500 Greatest Songs of All Time}.
In contrast, Koops et al. \cite{koops2019subjectivity} report an inter-annotator agreement rate of $76\%$ for the root note on four different annotations of a 50-song subset of the Billboard dataset \cite{burgoyne2011billboard}. 
To address annotation subjectivity, Koops et al.~\cite{koops2017personalization, koops2019subjectivity} propose a personalised chord estimation framework that adapts labels to individual annotator vocabularies. 
Their method computes Shared Harmonic Interval Profiles (SHIPs) from multiple reference annotations aligned with CQT frames and trains a neural network to predict user-specific chord labels, offering an alternative to fixed-vocabulary systems.
While this approach offers valuable insights into annotation variability, it addresses personalization rather than resolving fundamental inter-annotator disagreement. In contrast, our proposed method develops generalized harmonic representations grounded in music theory principles, thereby eliminating dependence on predefined chord vocabularies. 

Moreover, our method applies Label Smoothing (LS), a technique employed to enhance the generalisation and learning speed of multi-class neural networks. 
Originally proposed in \cite{szegedy2016fundationalls}, LS redistributes a portion of the probability mass from the observed class to other classes, thereby softening the distribution and generating what is referred to as \emph{soft targets}. 
This regularisation method has found widespread application in various state-of-the-art models across domains such as image classification, language translation, and speech recognition. 
\mr{It} has also been tested for music classification tasks \cite{buisson2033classificationsmoothing}, \mr{improving performance and reducing overfitting in small network training.}

While LS primarily serves as a regularisation technique, numerous studies have delved into its potential for encoding meaningful relationships among different categories. 
For instance, in \cite{liu2021similaritysmoothing}, authors propose an impactful method for generating more reliable soft labels that explicitly consider the relationships among various categories. 
Similarly, in \cite{lienen2021relaxzation}, a novel approach known as \emph{label relaxation} is introduced, which involves replacing a degenerate probability distribution associated with an observed class label, not by a single smoothed distribution but rather by a larger set of candidate distributions.

We integrate label smoothing into a model based on the conformer architecture~\cite{gulati2020conformer}, which has recently emerged in Automatic Speech Recognition (ASR) as an effective way of modelling global and local audio dependencies by leveraging a combination of CNNs and Transformer architectures. 
It has showcased remarkable success across various tasks not only in speech \cite{chiu2022selfsupervised} but also in music \cite{won2023foundation}, including melodic transcription \cite{tamer2023violin}, representation learning \cite{duong2022conformerrepresentation}, and music audio enhancement \cite{chae2023conformerenhancement}.
\mr{It also proved to be suitable for harmonic analysis, as it has been used for audio–chord alignment~\cite{Poltronieri2024ChordSync} and more recently adapted for chord estimation~\cite{akram2025chordformer}, where it is combined with the large-vocabulary decoding scheme proposed in~\cite{jiang2019large-vocabulary-decomposition}.}

\section{Methods}
\label{sec:methods}


\mr{We present a four-part investigation into chord estimation:}
\begin{enumerate}[label=(\roman*)]
  \item we conduct a comprehensive analysis of inter-annotator agreement across multiple chord similarity metrics, assessing how non-binary metrics measure inter-annotator agreement scores;
  \item we introduce a new perceptually-informed distance metrics and we demonstrate how it can improve agreement between annotators;
  \item we introduce a consonance-based label smoothing that leverages consonance to improve chord recognition; 
  \item we present a novel chord label encoding/decoding methodology, inspired by \cite{mcfee2017cnn}.
\end{enumerate}

\subsection{Analysis of Inter-Annotator Agreement}
\label{sec:agreement}

As outlined in Section \ref{sec:intro}, standard metrics employed to evaluate chord estimation systems have traditionally relied on binary comparison approaches \cite{pauwels2019twentyyears}. 
The most fundamental of these is the binary distance \( B_{\text{dist}}(C_1, C_2) \), which is defined as 1 if \( C_1 = C_2 \), and 0 otherwise.

When evaluating chord annotations or estimation algorithms, this binary comparison is typically weighted by the duration of each chord segment to compute the Chord Symbol Recall ($CSR$) \cite{harte2010towards}:
\begin{equation}
    \label{eq:csr}
    CSR = \dfrac{\mid S_a \cap S_e \mid}{\mid S_a \mid}.
\end{equation}
where $S_e$ represents the set of time segments where the estimated chords match the reference annotations, and $S_a$ represents the total duration of annotated segments.




\ap{In addition to overall binary agreement, several granular evaluation metrics have been introduced, each capturing different levels of harmonic detail. 
The \textit{Root} metric compares only the root note, ignoring chord quality and extensions. 
\textit{Thirds} extends this by incorporating major and minor third intervals. 
\textit{Triads} evaluate the full triadic structure—including major, minor, augmented, diminished, and suspended chords, up to the fifth scale degree. 
\textit{Tetrads} consider closed-voicing chords with extended tones (e.g., 9ths, 11ths, 13ths) collapsed into a single octave. 
The \textit{Sevenths} metric restricts evaluation to a predefined set of common seventh chord types. 
Finally, the \textit{MIREX} metric deems an estimate correct if it shares at least three pitch classes with the reference chord, regardless of root or quality. 
These metrics can optionally account for chord inversions by requiring the bass note to match as well. 
All are implemented in the \texttt{mir\_eval} library~\cite{raffel2014mireval}, which is the de facto standard for chord estimation evaluation.}
These metrics have been consistently used in literature to assess inter-annotator agreement in chord datasets, reporting agreement rates for the root note ranging from $76\%$~\cite{koops2019subjectivity} to $92\%$~\cite{clercq2011rockcorpus}.

\ap{However, to overcome the inherent limitations of binary evaluation metrics, recent research has introduced alternative measures. McLeod et al.~\cite{mcleod2023newmetrics} proposed three new metrics that more accurately represent musical relationships among chords: Spectral Pitch Similarity, Tone-by-Tone Distance, and Mechanical Distance.}

\ap{\textit{Spectral Pitch Similarity}, which assesses perceived pitch content based on psychoacoustic principles, lies beyond the scope of this study. 
On the other hand, \textit{Tone-by-Tone Distance} (TbT) treats chords as pitch-class sets, categorising pitches as either tonal or neutral. 
This metric quantifies chord similarity by measuring the proportion of shared pitch classes, resulting in a distance value reflecting their pitch-content similarity.
In contrast, \textit{Mechanical Distance} provides a more granular evaluation by approximating the physical distance between chord labels as they would be played on an instrument. 
It extends Tone-by-Tone Distance by quantifying not only the proportion of incorrect pitches but also the magnitude of each deviation from the target chord, by default measured in semitones.}

\begin{table}[b!]
\centering
\begin{tabular}{@{}l|cccc@{}}
\toprule
& \multicolumn{4}{c}{\textbf{CASD Dataset}} \\
\textbf{Metric} & \textbf{mir\_eval}$\uparrow$ & \textbf{TbT}$\uparrow$ & \textbf{Mech}$\downarrow$ & \textbf{Mech-Cons}$\downarrow$ \\
\midrule
Root & 0.757 & 0.773 & 0.817 & 0.604  \\
Thirds & 0.741 & 0.773  & 0.896 & 0.716 \\
Triads & 0.710 & 0.796  & 1.549 & 1.663 \\
MajMin & 0.734 & 0.803 & 1.465 & 1.577 \\
Tetrads & 0.572 & 0.786 & 1.859 & 1.803 \\
Sevenths & 0.592 & 0.794 & 1.771 & 1.715 \\
MIREX & 0.744 & 0.786 & 1.859 & 1.803 \\
\midrule
& \multicolumn{4}{c}{\textbf{Random Dataset}} \\
\textbf{Metric} & \textbf{mir\_eval}$\uparrow$ & \textbf{TbT}$\uparrow$ & \textbf{Mech}$\downarrow$ & \textbf{Mech-Cons}$\downarrow$ \\
\midrule
Root & 0.145 & 0.158 & 2.914 & 2.336 \\
Thirds & 0.140 & 0.158 & 2.914 & 2.336 \\
Triads & 0.121 & 0.253  & 5.536	& 5.861 \\
MajMin & 0.124 & 0.248 & 5.530	& 5.958 \\
Tetrads & 0.121 & 0.253 & 5.536 & 5.861 \\
Sevenths & 0.124 & 0.248 & 5.530 & 5.961 \\
MIREX & 0.121 & 0.253 & 5.536 & 5.861 \\
\bottomrule
\end{tabular}
\caption{Inter-Annotator Agreement Scores for Chord Annotations. TbT = Tone-by-Tone distance, Mech = Mechanical distance, Mech-Cons = Mechanical with Consonance distance. 
}
\label{tab:chord-agreement}
\end{table}

\begin{figure*}[ht]
    \centering
    \includegraphics[width=1\linewidth]{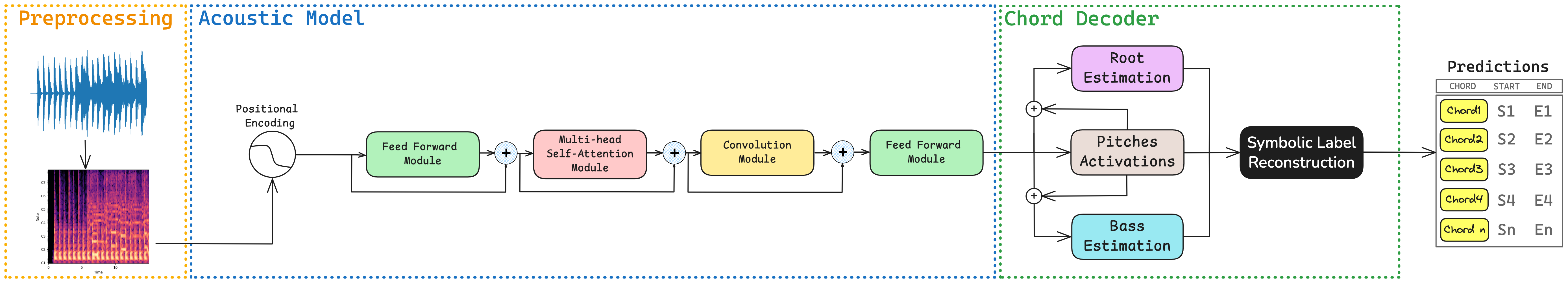}
    \caption{Overview of the Conformer model architecture, which comprises the preprocessing stage, the conformer-based model, and the symbolic chord decoder.}
    \label{fig:architecture}
\end{figure*}

\ap{While this approach introduces a more musically grounded notion of distance, the original formulation of Mechanical Distance still treats all semitone deviations as perceptually equivalent. This simplification overlooks the fact that, in Western tonal harmony, the perceptual impact of an interval depends not only on its size but also on its harmonic function.} To address this limitation, we propose an extension that incorporates consonance-based weighting into the Mechanical Distance. Specifically, we introduce the Mechanical-Consonance metric, which integrates the perceptual consonance vector presented in~\cite{giannos2021gct}, grounded in empirical studies of Western tonal harmony. 

The consonance vector is defined as:

\begin{equation}
\label{eq:consonance-vector}
vt = [0, 7, 5, 1, 1, 2, 3, 1, 2, 2, 4, 6]
\end{equation}
where each position corresponds to an interval in semitones, assigning lower values to more consonant intervals. 
For instance, perfect fifths and thirds (P5, m3, M3) receive the lowest score ($1$), indicating high consonance, while dissonant intervals such as major sevenths, minor seconds, and tritones are assigned higher values (up to $7$). 
Intervals of intermediate consonance, such as fourths and sixths, are assigned moderate values. 
By weighting semitone deviations using this vector, the Mechanical-Consonance metric adjusts the contribution of each error based on its perceptual salience. 

\ap{As a first contribution of this paper, we assess inter-annotator agreement across various chord granularity levels (e.g., root, thirds, triads) by comparing standard \texttt{mir\_eval} metrics with Tone-by-Tone Distance and Mechanical Distance. To align these non-binary metrics with the granularity levels typically employed in ACE evaluations, we apply two heuristics: 
\begin{enumerate*}[label=(\roman*)]
  \item restricting comparisons to the pitch ranges considered by the respective \texttt{mir\_eval} metrics (e.g., pitches up to the fifth of the chord for the \textit{MajMin} metric); and
  \item limiting comparisons only to chords included in the \texttt{mir\_eval} metric evaluation (e.g., diminished and seventh chords are excluded from the \textit{MajMin} metric).
\end{enumerate*}}

We conduct this analysis on the Chordify Annotator Subjectivity Dataset (CASD)~\cite{chordify2019}, which represents the largest available dataset for assessing chord annotation agreement and was previously used for similar studies~\cite{koops2019subjectivity}.

Moreover, to establish baseline performance and assess metric reliability, we conduct parallel experiments on a synthetically generated dataset replicating CASD's structure ($50$ tracks with $4$ annotations each), but populated with randomly generated chord sequences that preserve both \mr{its} 
chord vocabulary and sequence-length distributions. 

Table~\ref{tab:chord-agreement} reports the results for both the CASD and synthetic datasets, highlighting the performance and reliability of each metric across different evaluation settings.
\ap{To aid interpretation, we first clarify the nature and scaling of each metric under comparison.}

\ap{The \texttt{mir\_eval} metrics are formulated as similarity measures, returning values in the range $[0,1]$, where 1 indicates perfect agreement and 0 indicates complete disagreement. In contrast, Tone-by-Tone Distance is defined as a distance metric in $[0,1]$, with 0 indicating identical pitch-class content and 1 indicating no overlap; we convert it to a similarity score by computing $1 - \texttt{TbT}$. Mechanical Distance returns an unbounded distance value influenced by the number of notes in the chords, the sequence length, and the underlying pitch distance function. Due to these variable factors, we report Mechanical Distance in its original form without normalisation, as any fixed rescaling would obscure meaningful differences.}

The results show a clear separation between the CASD and random datasets, confirming that all metrics are sensitive to musically meaningful agreement.
\ap{TbT similarity scores are remarkably stable across all chord granularity levels, including more complex ones such as Sevenths and Tetrads. 
In the random dataset, TbT returns consistently higher values than \texttt{mir\_eval}, and scores increase progressively as more notes are considered in the evaluation (e.g., from Root to Sevenths). 
This trend indicates that TbT is more permissive than discrete match-based approaches and more sensitive to coincidental pitch-class overlap when more components are involved.}

\ap{Mechanical Distance exhibits lower agreement for simpler structures (e.g., Root and Thirds), closely mirroring the \texttt{mir\_eval} pattern. This is also reflected in the random dataset, where increasing the chord complexity leads to proportionally larger distances.}

\ap{Mechanical-Consonance generally produces lower scores for the CASD dataset and higher scores for the random dataset compared to its unweighted counterpart. Notably, the mean difference between CASD and random results is $3.326$ for Mechanical Distance and $3.471$ for Mechanical-Consonance. 
This larger separation supports the idea that inter-annotator disagreements are not random but often occur between harmonically related chords. 
The consonance-weighted formulation reinforces this insight by penalising perceptually dissonant deviations more heavily, further distinguishing musically plausible disagreements from unstructured noise.}

\begin{figure*}[ht]
    \centering
    \includegraphics[width=.7\linewidth]{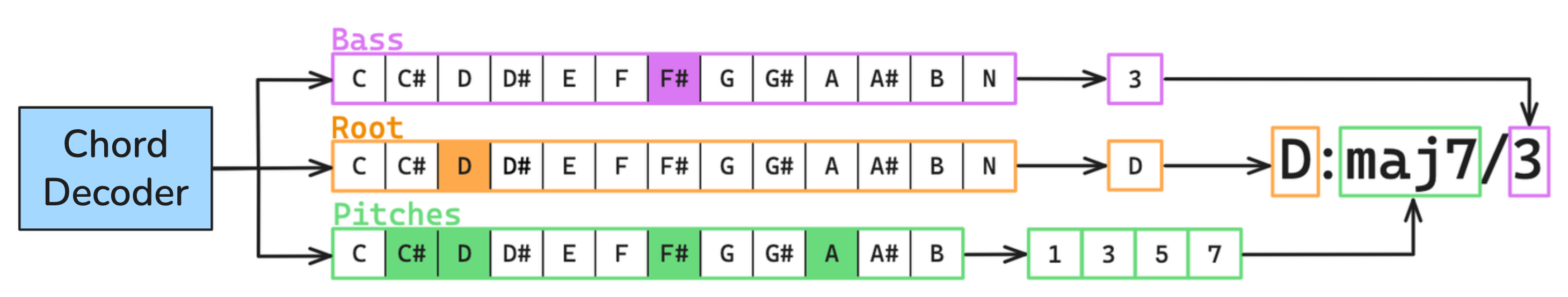}
    \caption{Example of chord label decoding for a \texttt{D:maj7/3} chord using the decomposed decoder, inspired by \cite{mcfee2017cnn}.}
    \label{fig:chord-decoder}
\end{figure*}

\subsection{Proposed Model}

As a second contribution, this paper presents a novel ACE model, illustrated in Figure~\ref{fig:architecture}, which leverages the Conformer architecture~\cite{gulati2020conformer}.
As a first step, the audio is first resampled to a sampling rate of $22050$ Hz, and a hop size of $2048$ is applied.
Then, the Constant-Q Transform (CQT) features are calculated on $6$ octaves starting from $C1$, with $24$ bins per octave, resulting in a total of $144$ bins.
The CQT features are fed to a conformer encoder~\cite{gulati2020conformer} before being passed to the decoder layers.

\subsubsection{Chord Decomposition and Decoding}
\label{subsec:vocabulary}

Label encoding follows a similar approach as \cite{mcfee2017cnn}. 
\mr{Root and bass notes are encoded as a 13-dimensional one-hot vector, where the first $12$ positions represent the semitones from $C$ to $B$, and the last one indicates silence (denoted as $N$).}
Chord tones are encoded using a 12-dimensional multi-hot vector, where each dimension indicates the presence ($1$) or absence ($0$) of a pitch class in the chord.



\bug{The output of the Conformer layers is first passed through a fully connected head to predict chord tones. 
These chord predictions then serve as conditioning information for two additional components: bass and root prediction. 
Each of these components employs a feature fusion mechanism that concatenates the original Conformer features with the chord logits, creating an enriched representation that captures both the acoustic context and the predicted harmonic content. 
This hierarchical approach reflects the musical intuition that bass and root notes are contextually dependent on the overall harmonic content, rather than treating all three components as independent prediction tasks.
}
To train the model, we use a composite loss that aligns with this encoding scheme. Cross-entropy loss is applied to root and bass predictions, and binary cross-entropy loss is used for chord tone predictions. Additionally, we introduce a regularisation term that penalises discrepancies between the predicted and actual number of active pitch classes.

\ap{The total loss is defined as:
\begin{equation}
    \begin{split}
        \mathcal{L} &= \lambda_{\text{root}} \, \mathcal{L}_{\text{CE}}^{\text{root}} + \lambda_{\text{bass}} \, \mathcal{L}_{\text{CE}}^{\text{bass}} + \lambda_{\text{chord}} \, \mathcal{L}_{\text{BCE}}^{\text{chord}} \\
        &\quad\quad\quad\quad\quad  + \lambda_{\text{card}} \, \lVert \hat{c} - c \rVert_1
    \end{split}
\end{equation}
}
\mr{where $c$ and $\hat{c}$ are the number of active notes in the ground truth and those predicted above a threshold, respectively.}

\ap{Differently from~\cite{mcfee2017cnn}, where the outputs of the bass, root, and pitch activation predictions are combined and passed through a final linear layer to predict chord labels, we directly use these three components to reconstruct the final chord label. The novelty of this approach lies in the fact that, unlike vocabulary-constrained decoding strategies such as~\cite{jiang2019large-vocabulary-decomposition}, our method does not require a predefined chord vocabulary.}

\ap{Chord labels are reconstructed from the predicted probabilities in a modular decoding process.} First, the root note is identified by selecting the pitch class with the highest predicted probability, which is then mapped to its symbolic representation. For the chord tones, a fixed threshold (default: 0.5) is applied to the predicted pitch activations; only pitches exceeding this threshold are retained. These pitch classes are then converted into intervals relative to the predicted root. An analogous procedure is applied to the bass prediction, allowing the full reconstruction of the chord structure, as illustrated in Figure~\ref{fig:chord-decoder}. 
Finally, the decoded chord is passed to the \texttt{harte\_library}\footnote{\url{https://github.com/andreamust/harte-library}}, which implements utilities for converting the predicted chord label into the respective shorthand notation.

\begin{table*}[ht!]
\centering
\resizebox{\textwidth}{!}{%
\begin{tabular}{lll|ccccccc|ccc}
\toprule
  \textbf{Model} &
  \textbf{Vocab} &
  \textbf{Smooth} &
  \textbf{Root$\uparrow$} &
  \textbf{MajMin$\uparrow$} &
  \textbf{Thirds$\uparrow$} &
  \textbf{Triads$\uparrow$} &
  \textbf{Tetrads$\uparrow$} &
  \textbf{7th$\uparrow$} &
  \textbf{MIREX$\uparrow$} &
  \textbf{TbT$\uparrow$} &
  \textbf{Mech$\downarrow$} &
  \textbf{MechCons$\downarrow$} \\ \midrule

Ours & 170 & -  & 81.4 & 77.5 & 78.1 & 72.3 & 59.6 & 64.7 & 79.4 & 77.9 & 1.55 & 1.35 \\ 
Ours & Decom. & -  & 83.4 & 77.2 & 79.7 & 72.2 & 59.2 & 64.6 & 79.3 & 80.5 & 1.57 & 1.37 \\ 
Ours & Decom. & Cons.  & \textbf{84.0} & \textbf{77.8} & \textbf{80.3} & \textbf{72.7} & \textbf{60.8} & \textbf{66.0} & \textbf{79.8} & \textbf{81.7} & \textbf{1.44} & \textbf{1.30} \\ 
\midrule
BTC & 170 & -  & 81.6 & 77.3 & 78.4 & 72.1 & 60.0 & 65.7 & 79.0 & 78.4 & 1.60 & 1.40 \\ 
BTC & Decom. & -  & 82.9 & 76.0 & 79.2 & 70.9 & 57.2 & 62.4 & 77.4 & 80.4 & 1.52 & 1.35 \\ 
BTC & Decom. & Cons.  & 82.8 & 76.1 & 79.3 & 70.9 & 59.5 & 64.7 & 79.0 & 80.7 & 1.49 & 1.32 \\ 
\bottomrule
\end{tabular}%
}
\caption{Performance comparison across different model variants using both standard \texttt{mir\_eval} metrics and non-binary metrics. Results are reported for our conformer-based model with and without the decomposition decoder and consonance-based label smoothing. Additionally, we compare these settings with the BTC model~\cite{park2019bi-directionaltransformer}.}
\label{tab:results}
\end{table*}

\subsubsection{Consonance-based Smoothing}
\label{subsec:smoothing}

We introduce a novel label smoothing technique that leverages music-perceptual knowledge by incorporating consonance relationships between pitch classes. 
Unlike conventional label smoothing that uniformly distributes probability mass across incorrect classes, our approach allocates probability according to the consonance relationship between pitch classes.

Let $\mathbf{c} = [c_0, c_1, \ldots, c_{11}] \in \mathbb{R}^{12}$ be a consonance vector where each element $c_i$ quantifies the dissonance level of the interval $i$ semitones above the reference pitch. 
Lower values of $c_i$ indicate more consonant intervals (e.g., perfect fifth, major third). We transform this vector into a similarity measure $\mathbf{s} \in \mathbb{R}^{12}$ as follows:

\begin{equation}
\mathbf{s} = 1 - \frac{\mathbf{c}}{\max(\mathbf{c})}
\end{equation}

This ensures that more consonant intervals receive higher similarity scores, with perfect consonance (unison) having a similarity of 1.
For a given target pitch class $t \in \{0, 1, \ldots, 11\}$ and smoothing factor $\alpha \in [0, 1]$, we define the smoothed target distribution $\mathbf{q} \in \mathbb{R}^{12}$ as:

\begin{equation}
q_i = 
\begin{cases} 
1 - \alpha & \text{if } i = t \\
\alpha \cdot s_{(i-t) \bmod 12} & \text{if } i \neq t
\end{cases}
\end{equation}
The distribution is then normalised to ensure $\sum_{i=0}^{11} q_i = 1$:

\begin{equation}
\mathbf{q} = \frac{\mathbf{q}}{\sum_{i=0}^{11} q_i}
\end{equation}

This formulation creates a probability distribution where the target class $t$ receives the highest probability $(1-\alpha)$, while the remaining probability mass $\alpha$ is distributed among other pitch classes proportionally to their consonance relationship with the target. For example, when the true class is C (0), pitch classes G (7) and F (5) will receive higher probability than more dissonant intervals like C\# (1) or B (11), reflecting their stronger harmonic relationships.



\section{Evaluation}
\label{sec:evaluation}

In this section, compare the performance of the proposed ACE model with a state-of-the-art method~\cite{park2019bi-directionaltransformer}, using standard \texttt{mir\_eval} metrics, Tone-by-Tone (TbT) similarity, and Mechanical distances. 
Additionally, we evaluate the effectiveness of the proposed chord decoder by benchmarking it against a conventional frame-wise classification approach, focusing on its ability to accurately capture chord inversions using the inverted \texttt{mir\_eval} metrics.


\bug{All chord annotations were sourced from ChoCo~\cite{deberardinis2023choco}, which provides standardized labels in Harte syntax~\cite{harte2005chord}. Specifically, we use annotations from the Isophonics dataset~\cite{mauch2009beatles} and the McGill Billboard corpus~\cite{burgoyne2011expert} for training and validation, while the RWC Pop~\cite{goto2002rwc} and USPop datasets~\cite{jiang2019large-vocabulary-decomposition} serve as test sets.
This setup enables evaluation of both model performance and generalization across diverse chord vocabularies.}

To increase data density while preserving local harmonic continuity, each track is segmented into 20-second excerpts with $50\%$ overlap. 
We employ data augmentation by transposing both audio and targets from $-5$ to $+6$ semitones.
During training, we use the \emph{AdamW} optimiser and cosine annealing learning rate scheduler to dynamically adjust the learning rate during training cycles. 
Additionally, we adopted mixed precision training \cite{paulius2017mixedprecision} to accelerate training. 
To prevent overfitting, we implement early stopping, terminating training when performance on a validation set ceased to improve after $10$ epochs. 
The code and all hyper-parameters used in the experiments are available on the GitHub repository of the project\footnote{\url{https://github.com/andreamust/consonance-ACE}}.

\begin{table}[h!]
\centering
\resizebox{.96\columnwidth}{!}{%
\begin{tabular}{l|cccc}
\toprule
\textbf{Metric} & \textbf{BTC} & \textbf{Ours} & \textbf{Ours} & \textbf{Ours} \\
\midrule
Vocab. & 170 & 170 & Decom. & Decom. Cons. \\
\midrule
MajMin Inv.$\uparrow$ & 71.5 & 72.4 & \textbf{75.6} & \textbf{75.6} \\
Thirds Inv.$\uparrow$ & 72.6 & 72.9 & 77.2 & \textbf{77.9} \\
Triads Inv.$\uparrow$ & 67.2 & 67.6 & 70.2 & \textbf{70.8} \\
Tetrads Inv.$\uparrow$ & 56.2 & 55.7 & 57.7 & \textbf{59.4} \\
Sevenths Inv.$\uparrow$ & 60.8  & 60.0 & 62.9 & \textbf{64.4} \\
\bottomrule
\end{tabular}%
}
\caption{Performance comparison on inverted chords between traditional architectures and the proposed decomposed model, evaluated using \texttt{mir\_eval} metrics.}
\label{tab:results_inverted}
\end{table}

\subsection{Evaluation of the ACE Model}

We evaluate our model using TbT similarity, Mechanical Distance, and its consonance-weighted variant, as introduced in Section~\ref{sec:agreement}, alongside standard binary metrics from \texttt{mir\_eval}~\cite{raffel2014mireval}. 
For comparison, we adopt the BTC model~\cite{park2019bi-directionaltransformer}, a state-of-the-art baseline for audio chord estimation. 
We reimplemented and retrained the BTC model using the hyper-parameter settings specified in the original paper, enabling a direct evaluation of our proposed decomposition-based decoder and the impact of consonance-informed label smoothing.
The experimental results are summarized in Table~\ref{tab:results}.

As noted by~\cite{jiang2019large-vocabulary-decomposition}, differences among models are often marginal when evaluated with standard metrics. 
This holds true in our comparison: both models yield similar results on the standard classification task over a 170-class chord vocabulary. 
However, our proposed decomposition-based decoder consistently outperforms the standard frame-wise classification architecture across several metrics, with the advantage of not relying on a fixed chord vocabulary. 
Notably, we observe the greatest improvement in Root and Thirds metrics.
Additionally, the use of non-binary metrics further highlights the benefits of the proposed decoder.
As shown in Table~\ref{tab:results_inverted}, inverted metrics also improve when using the proposed decomposed decoder. 
\bug{
This improvement stems from the fact that the proposed chord decoding scheme explicitly predicts the bass note, enabling accurate inversion prediction--a capability that standard chord classification approaches inherently lack.
The same trend is confirmed when applying the decomposed decoder to the BTC model, which yields performance increases across several metrics, especially the non-binary ones.}

When integrating the consonance-weighted loss on root and bass predictions within the proposed decoding architecture, performance slight improvement on all metrics. 
Notably, improvements are observed also on non-binary metrics and on inverted chords. The trend is also confirmed when applying consonance smoothing to the BTC model with the decomposed decoder.
\bug{Overall, evaluation results suggest that both the proposed decoder and the consonance smoothing improve accuracy in most metrics, and led to predictions more consonant to the target.}

\section{Conclusions}
\label{sec:conclusions}

In this paper, we presented a novel model for Audio Chord Estimation based on the conformer architecture, enhanced with a consonance-informed label smoothing strategy and a decomposition-based decoding scheme. 
The motivation for incorporating perceptual smoothing emerged from our inter-annotator agreement analysis, which employed non-binary distance metrics and revealed that annotation discrepancies often involve harmonically related chords. Building on these insights, we introduced a learning strategy that integrates consonance-weighted targets into the training process. 

Experimental results show that the proposed model achieves strong performance across both standard and non-binary evaluation metrics, with notable gains in capturing fine-grained harmonic relationships. 
Additionally, the proposed decomposition decoder not only enables chord prediction without relying on a fixed chord vocabulary, but also
contributes to consistent performance improvements.







\section{Acknowledgments}

This work is supported by IA y Música: Cátedra en Inteligencia Artificial y Música (TSI-100929-2023-1), funded by the Secretaría de Estado de Digitalización e Inteligencia Artificial, and the European Union-Next Generation EU, under the program Cátedras ENIA 2022 para la creación de cátedras universidad-empresa en IA, and IMPA: Multimodal AI for Audio Processing (PID2023-152250OB-I00), funded by the Ministry of Science, Innovation and Universities of the Spanish Government, the Agencia Estatal de Investigación (AEI) and co-financed by the European Union.




\bibliography{bibliography}

\end{document}